\documentclass[aps,prl,amsmath,amssymb,twocolumn,floatfix]{revtex4}
\usepackage{graphicx}
\usepackage{bm}
\usepackage{epsfig}

\newcommand{\p}{{\rm poly}}

\newcommand{\cl}{{\rm c}}
\newcommand{\sh}{{\rm s}}
\newcommand{\eps}{\epsilon}
\newcommand{\drc}{\delta\rho_\cl}
\newcommand{\drs}{\delta\rho_\sh}
\newcommand{\dpc}{\delta\phi_\cl}
\newcommand{\dps}{\delta\phi_\sh}
\newcommand{\T}{^{\text T}}
\newcommand{\Mmat}{\bm{M}}
\newcommand{\delvec}{\bm{\delta}}
\newcommand{\vvec}{\bm{v}}
\newcommand{\evec}{\bm{e}}

\newcommand{\order}{\mathcal{O}}
\newcommand{\half}{\frac{1}{2}}

\newcommand{\be}{\begin{equation}}
\newcommand{\ee}{\end{equation}}
\newcommand{\bea}{\begin{eqnarray}}
\newcommand{\eea}{\end{eqnarray}}
\newcommand{\beastar}{\begin{eqnarray*}}
\newcommand{\eeastar}{\end{eqnarray*}}
\newcommand{\eq}[1]{~(\ref{#1})}

\begin{document}

\title{Weakly polydisperse systems: Perturbative phase diagrams that
  include the critical region}

\author{Peter Sollich}
\affiliation{King's College London, Department of Mathematics, Strand,
London WC2R 2LS, United Kingdom.}

\date{\today}

\begin{abstract}The phase behaviour of a weakly polydisperse system,
  such as a colloid with a small spread of particle sizes, can be
  related perturbatively to that of its monodisperse counterpart. I
  show how this approach can be generalized to remain well-behaved near
  critical points, avoiding the divergences of existing
  methods and giving access to some of the key qualitative features of
  polydisperse phase equilibria. The analysis explains also why in purely
  size-polydisperse systems the critical point is, unusually, located
  very near the maximum of the cloud and shadow curves.

%
%64.10.+h General theory of equations of state and phase equilibria
%82.70.-y Disperse systems; complex fluids
%05.20.-y Classical statistical mechanics
%68.35.Rh Phase transitions and critical phenomena
%
%64.75.-g Phase equilibria (part of ``specific phase transitions'')
%61.25.H- Macromolecular and polymers solutions; polymer melts
%64.70.F- Liquid-vapor transitions
%
\end{abstract} 

\maketitle
\setcounter{totalnumber}{10}

Most soft materials are inherently polydisperse: spherical colloids
have a spread of diameters, polymers an effectively continuous
distribution of chain lengths etc. 
The
effects of polydispersity on equilibrium phase behaviour are both of
fundamental 
interest and of relevance to production, processing and shelf life 
of foodstuffs, personal hygiene products, paints and coatings etc.
Careful sample preparation can
reduce polydispersity, to e.g.\ diameter standard deviations of around
5\% for colloids~\cite{PusVan86}, but not eliminate it. 
Existing theories that calculate the effects of such weak
polydispersity by expanding around the
phase equilibria of a monodisperse reference
system~\cite{EvaFaiPoo98,Evans01} have proved useful, but break down
around critical points (CPs) where the relevant prefactors
scale with the compressibility.
%This is not due to the effects of non-classical critical behaviour and
%so occurs even within simple mean field theories. For the latter,
The critical region can be studied alternatively 
by a Landau expansion~\cite{RasCat03}; while not limited to weak
polydispersity, this is impractical for all but the simplest
models. Generic tools for predicting polydispersity
effects also in the critical region would clearly be useful: as we will
see, the qualitatively distinct features of
polydisperse phase behaviour manifest themselves most
clearly here, and critical parameters can be estimated reliably from
simulations~\cite{WilSol04,WilFasSol04} and used to
constrain theoretical models.
My aim here is to supply such a tool, by formulating a
perturbative approach that remains applicable in the critical
region. The approach allows for the prediction of the full phase behaviour of
generic weakly polydisperse materials from that of the monodisperse
reference system. As a by-product, the analysis rationalizes
previously unexplained peculiarities in the phase behaviour
of purely size-polydisperse system. The key idea is that, because a
polydisperse system has a critical temperature different from its
monodisperse counterpart, one has to allow perturbations not just of
the coexisting densities but also of temperature.

Denote the polydisperse attribute, e.g.\ particle diameter, by
$\sigma$, and its average in the parent phase that is
allowed to phase separate by $\sigma_0$. In terms of the scaled
deviations from the parental mean, $\eps=(\sigma-\sigma_0)/\sigma_0$,
the normalized parent size distribution $p(\eps)$ then has zero mean
and, by the assumption of weak polydispersity, small variance
$w=\int\!d\eps\,p(\eps)\eps^2$. We describe a generic phase of the
system by its (number) density distribution $\rho(\eps)$. Its excess
Helmholtz free energy density in units of $k_{\rm B}T$, $\tilde f_\p$, is a
functional of $\rho(\eps)$ and an ordinary function of inverse
temperature $\beta=1/k_{\rm B}T$. For weak polydispersity this can be
expanded up to $\order(\eps^2)$ as~\cite{EvaFaiPoo98,Evans01}:
\be
\tilde f_\p([\rho(\eps)],\beta) = \tilde f+A\rho_1+B\rho_2+C\rho_1^2
\label{f_expansion}
\ee
Here the monodisperse reference free energy $\tilde f$ is a function
of the overall number density $\rho_0=\int\!d\eps\,\rho(\eps)$ and
$\beta$, as are the coefficients $A$, $B$, $C$; we abbreviate
$\rho\equiv\rho_0$ below. The higher-order moment densities are
defined as $\rho_{n}=\int\!d\eps\,\rho(\eps)\eps^n$ ($n=1,2$). From
$\tilde f_\p$ one finds the excess chemical potentials $\tilde\mu(\eps)
=\delta \tilde f_\p/\delta \rho(\eps)$
\[
\tilde\mu(\eps) = \tilde f_\rho + A_\rho \rho_1 + B_\rho \rho_2 +
C_\rho \rho_1^2
+ (A+2C\rho_1)\eps + B\eps^2
\]
where subscripts indicate derivatives, i.e.\ $\tilde f_\rho \equiv \partial
\tilde f/\partial\rho$ etc. Equality of the full chemical potentials
$\mu(\eps)=\ln\rho(\eps)+\tilde\mu(\eps)$ then implies that phases in
coexistence have density distributions differing by a factor of the form
$\exp(\lambda_0 + \lambda_1\eps + \lambda_2\eps^2)$, as expected from
the moment structure of\eq{f_expansion}~\cite{SolCat98,SolWarCat01}.
The pressure $\Pi_\p = \rho + \int\!d\eps\,\rho(\eps)\tilde\mu(\eps) -
\tilde f_\p$ becomes
\bea
\Pi_\p = \Pi + A_\rho \rho\rho_1 + B_\rho \rho\rho_2 +
(C_\rho \rho+C)\rho_1^2
\eea
with the monodisperse reference pressure $\Pi=\rho
+ \rho\tilde f_\rho-\tilde f$.

In a polydisperse system the monodisperse binodal splits into cloud
and shadow curves~\cite{Sollich02}. The former records the density of
the majority phase at the onset of coexistence, the latter that of the
incipient phase. The cloud phase is then identical to the parent, with a
density distribution of the form $\rho\, p(\eps)$, while that of the
coexisting shadow can be written $\rho\, p(\eps)\exp(\lambda_0 +
\lambda_1\eps + \lambda_2\eps^2)$. Equality of pressures and of
the $\order(\eps^0)$, $\order(\eps^1)$ and $\order(\eps^2)$ terms in
the chemical potentials $\mu(\eps)$ gives the conditions
\beastar
\Pi + w B_\rho \rho^2 &=& \Pi' + A_\rho' \rho'\rho_1' + B_\rho' \rho'\rho_2' +
(C_\rho' \rho'+C')\rho_1'{}^2
\\
\tilde f_\rho + wB_\rho \rho &=&
\tilde f_\rho' + A_\rho' \rho_1' + B_\rho' \rho_2' + C_\rho'
\rho_1'{}^2 + \lambda_0
\\
A &=& A'+2C'\rho_1' + \lambda_1
\\
B &=& B' + \lambda_2
\eeastar
where unprimed/primed quantities relate to the cloud and shadow phase,
respectively, e.g.\ $\rho'$, $\rho_1'$, $\rho_2'$ are density moments
in the shadow and $A_\rho\equiv A_\rho(\rho,\beta)$, $A_\rho'\equiv
A_\rho(\rho',\beta)$; we have used that
$\rho_1=\rho\langle\eps\rangle=0$ and
$\rho_2=\rho\langle\eps^2\rangle=w\rho$ in the cloud.

The basis of the perturbation expansion is the monodisperse phase
coexistence, with coexisting 
densities $\rho$, $\rho'=\rho\exp(\lambda_0)$ at temperature
$\beta$. One then has $w=\rho_1'=\rho_2'=0$ and
the Lagrange multipliers become $\lambda_0=\tilde f_\rho - \tilde
f_\rho' \equiv \Delta\tilde f_\rho$, $\lambda_1=A-A'\equiv \Delta A$,
$\lambda_2=B-B'\equiv\Delta B$. The $\Delta$-notation indicates
differences between coexisting monodisperse phases. When the parent is
made polydisperse, $\rho$ is perturbed to the cloud point density
$\rho_\cl=\rho+\drc$, and the Lagrange multipliers to
$\lambda_0=\Delta\tilde f_\rho + \delta\lambda_0$, $\lambda_1=\Delta A
+ \delta\lambda_1$, $\lambda_2=\Delta B + \delta\lambda_2$. Expanding
to linear order in all small quantities including $w$, the shadow
moments become $\rho_\sh'=\rho' + \drs'$ with
\be
\drs'=\rho'\{\drc/\rho+\delta\lambda_0+w[{\textstyle\half}
(\Delta A)^2+\Delta B]\}
\label{drs}
\ee
%
%or
%\[
%\delta\lambda_0 = \drs'/\rho' - \drc/\rho 
%- w[(\Delta A)^2/2+\Delta B]
%\]
%
and $\rho_1'=w\rho'\Delta A$, $\rho_2'=w\rho'$.  One now inserts these
into the four phase coexistence conditions above and expands, crucially also
allowing a temperature shift $\beta\to\beta+\delta\beta$. The fourth
conditions only serves to determine $\delta\lambda_2$
so can be omitted. The third condition reduces to
\be
A_\rho\drc - A_\rho'\drs' + \Delta A_\beta \,\delta\beta =
2 w \rho'C'\Delta A + \delta\lambda_1
\label{dlambda1_eq}
\ee
while the first two yield, after eliminating $\delta\lambda_0$ in
favour of $\drs'$ using\eq{drs},
%
%\beastar
%\Pi_\rho\drc + \Pi_\beta\delta\beta
% + w B_\rho \rho^2 &=& \Pi_\rho' \drs' +
% \Pi_\beta' \delta\beta + w A_\rho' \rho'^2\lambda_1 + w
% B_\rho' \rho'^2
%\\
%\tilde f_{\rho\rho}\drc +
%\tilde f_{\rho\beta}\delta\beta + w B_\rho \rho &=&
%\tilde f_{\rho\rho}' \drs' +\tilde f_{\rho\beta}'\delta\beta
% + w A_\rho' \rho_m'\lambda_1 + w B_\rho' \rho' + \delta\lambda_0
%\\
%A_\rho\drc + A_\beta\delta\beta &=& A_\rho'\drs'
%+A_\beta'\delta\beta + 2wC'\rho'\lambda_1 + \delta\lambda_1
%\\
%B_\rho \drc + B_\beta \delta\beta &=& B_\rho' \drs'
%+ B_\beta' \delta\beta + \delta\lambda_2
%\eeastar
%
\beastar
& &\!\!\!\!\!\!\Pi_\rho\drc - \Pi_\rho' \drs' + \Delta\Pi_\beta\,\delta\beta =
\nonumber\\
& &\ \ = w[A_\rho' \rho'^2\Delta A -\Delta(B_\rho \rho^2)]
\nonumber\\
& &\!\!\!\!\!\!(\tilde f_{\rho\rho}+1/\rho)\drc 
- (\tilde f_{\rho\rho}'+1/\rho') \drs' 
+ \Delta\tilde f_{\rho\beta}\, \delta\beta =
\nonumber\\
& &\ \ =
w\{A_\rho' \rho'\Delta A -  (\Delta A)^2/2 - \Delta(B_\rho \rho) - \Delta B\} 
\eeastar
The cofficients involving $\tilde f$ and $\Pi$ can be written in terms
of the full monodisperse free energy density including the ideal part,
$f=\rho(-1+\ln\rho)+\tilde f$: $\tilde f_{\rho\rho}+1/\rho =
f_{\rho\rho}$, $\tilde f_{\rho\beta}=f_{\rho\beta}$, $\Pi_\rho = \rho
f_{\rho\rho}$ and $\Pi_{\beta}=\rho f_{\rho\beta}-f_\beta$. Taking
appropriate linear combinations to eliminate $\drc$ or $\drs'$
then gives
%
%\beastar
%\rho f_{\rho\rho} \drc - \rho' f_{\rho\rho}' \drs'
%+ \Delta(\rho f_{\rho\beta}-f_\beta)\delta\beta
% &=& w[\rho'^2 A_\rho' \Delta A -\Delta(\rho^2 B_\rho)]
%\\
%f_{\rho\rho}\drc 
%- f_{\rho\rho}' \drs' 
%+ \Delta f_{\rho\beta}\, \delta\beta
%&=&
%w\{\rho' A_\rho'\Delta A -  {\textstyle\half}(\Delta A)^2 -
%\Delta(\rho B_\rho) - \Delta B\} 
%\\
%A_\rho\drc - A_\rho'\drs' + \Delta A_\beta \,\delta\beta &=& 
%2 w C'\rho'\Delta A + \delta\lambda_1
%\eeastar
%
%Multiply second equation by $-\rho$ and $-\rho'$ respectively and
%add first:
%
\bea
& &\!\!\!\!\!\!f_{\rho\rho}\drc 
+ [f_{\rho\beta} - \Delta f_\beta/\Delta\rho]\, \delta\beta =
\label{Mrow2}\\
& &\ \ = w \{\rho'[{\textstyle\half}(\Delta A)^2 +
\Delta B]/\Delta\rho - \rho B_\rho \}
\nonumber\\
& &\!\!\!\!\!\!f_{\rho\rho}' \drs'
+[f_{\rho\beta}' -\Delta f_\beta/\Delta\rho]\, \delta\beta =
\label{Mrow3}\\
& &\ \ = w\{\rho[{\textstyle\half}(\Delta A)^2 + \Delta B]/\Delta\rho
-\rho' B_\rho' -\rho'A_\rho' \Delta A \}
\nonumber
\eea
If temperature is held constant ($\delta\beta=0$), one retrieves
Evans' results~\cite{Evans01} for the cloud and shadow density shifts.
This approach breaks down at the CP because
$1/f_{\rho\rho}$ and $1/f_{\rho\rho}'$, which are proportional to
the compressibilities in the monodisperse reference phases, diverge.

In the more general setup here we can choose a prescription for
$\delta\lambda_1=\order(w)$ and deduce the temperature shift
$\delta\beta$, which will be of the same order. This freedom does not
mean that the perturbation theory is ill-defined. In fact, at a fixed
perturbed temperature $\beta_0=\beta+\delta\beta$ away from the
CP, the results are independent of $\delta\beta$.
To see this for e.g.\ the cloud point density, write\eq{Mrow2} as
\be
\rho_\cl(\beta_0)=\rho(\beta)+\rho_\beta(\beta)\delta\beta+w R(\beta)
\label{cloud}
\ee
The function $wR(\beta)$ is the r.h.s.\ of\eq{Mrow2} divided by
$f_{\rho\rho}$, with the $\beta$-dependence of the monodisperse
coexisting densities $\rho$ and $\rho'$ inserted. That the prefactor
of $\delta\beta$ must equal the derivative $\rho_\beta\equiv
d\rho/d\beta$ along the monodisperse binodal follows by setting $w=0$,
in which case there is no additional polydispersity-induced shift. Now
if $\delta\beta=0$~\cite{Evans01},
$\rho_\cl(\beta_0)=\rho(\beta_0)+w R(\beta_0)$.  This agrees to
$\order(w)$ with\eq{cloud}, for {\em any} temperature shift
$\delta\beta=\order(w)$, since
$\rho(\beta_0)=\rho(\beta)+\rho_\beta(\beta)\delta\beta+\order(w^2)$;
the difference between $R(\beta)$ and $R(\beta_0)$ likewise only gives
$\order(w^2)$ corrections.

\iffalse
Instead, we prescribe $\delta\lambda_1$ and deduce
$\delta\beta$. @@ Need to say something about why there is choice
here. Geometrically, choose a point on a shifted tangent to the
monodisperse binodal. Take point where $\beta+\delta\beta=\beta'$:
tangent at $\beta$. Take Evans' prescription evaluated at $\beta'$, no
temperature shift: tangent at $\beta'$. These two tangents are
``almost'' the same, and we're looking at their intersection with the
same horizontal (temperature $=\beta'$) line, so the density shifts
agree to $\order(w)$. Should check this mathematically? Set $w=0$, get
perturbation of $\rho$ and $\rho'$. @@
\fi
%
To guide the choice of $\delta\lambda_1$ one can use the intuition
that in a reliable perturbation theory corresponding monodisperse and
polydisperse state points should be mapped to each other. At the
CP, where cloud and shadow become identical, also their
size distributions do, so we must have $\lambda_1=\Delta A
+\delta\lambda_1=\delta\lambda_1=0$. This suggests taking
$\delta\lambda_1=0$ throughout the phase diagram; to eliminate
the function $C$ from\eq{dlambda1_eq}, however, we choose
\be
\delta\lambda_1= -2 w \rho'C'\Delta A
\label{dlambda1_choice1}
\ee
which still vanishes at the CP.

Equations (\ref{dlambda1_eq}--\ref{Mrow3}) with the
choice\eq{dlambda1_choice1} constitute our generalized perturbation
theory; the set of three linear equations is easily solved for $\drc$,
$\drs'$, $\delta\beta$. I show below that the resulting perturbations
do remain finite in the critical region for systems with
classical critical behaviour. This encompasses a large variety of
models used to describe polydisperse complex fluids, such as van der
Waals theory for liquids, free volume theory for colloid-polymer
mixtures, the Flory-Huggins and Wertheim's statistical associating
fluid theories for polymers etc (see references in~\cite{Sollich02}).
For monodisperse reference systems with non-classical critical
behaviour one expects that polydispersity will Fisher-renormalize the
critical exponents~\cite{KitDobYamNakKam97}, multiplying them by
$1/(1-\alpha)$. For small $w$ a crossover from the monodisperse to the
polydisperse exponents should then occur very close to the CP, and
this will not be accounted for by our perturbation theory. Studying
this crossover is an interesting issue for
future work. Because the specific heat exponent $\alpha\approx 0.11$
is small in $d=3$, however, 
the quantitative effects of Fisher renormalization may well be small
enough for the perturbative approach to remain an accurate
approximation.

To obtain an explicit expression for the CP shift due to
polydispersity, we start by assuming a smooth
(Landau-like) expansion of $f$ near criticality. The coexisting
densities then go as $\rho-\rho^*=-(\rho'-\rho^*)=
[2(\beta-\beta^*)/b]^{1/2}$ to leading order, with $b$ some
constant; asterisks denote quantities at the CP. Defining
$r=[2(\beta-\beta^*)/b]^{1/2}$, we can then write $\rho-\rho^*=r+ar^2$,
$\rho'-\rho^*=-r+ar^2$, $\beta=\beta^*+\frac{1}{2}br^2$ to $\order(r^2)$. The
coefficient $b$ is given by
$b=-f^*_{\rho\rho\rho\rho}/(3f^*_{\rho\rho\beta})$; $a$
will cancel to the order we require. To expand
our equations for small $r$ write\eq{dlambda1_eq}, together with
(\ref{Mrow2},\ref{Mrow3}) divided by $\Delta\rho$, in matrix form
$\Mmat\delvec=w\vvec$. Here $\delvec\T=(\drc,\drs',\delta\beta)$ and
$\vvec$ gathers the appropriate right-hand sides. Substituting
the above expansions for $\rho$, $\rho'$, $\beta$ and using
$f^*_{\rho\rho}=f^*_{\rho\rho\rho}=0$ at the CP then yields $\Mmat=\Mmat^*+\Mmat^*_r
r+\ldots$, $\vvec=\vvec^*+\vvec^*_r r + \ldots$ with
\bea
\Mmat^*=\left(\begin{array}{rrr}
A^*_\rho & -A^*_\rho & 0 \\
0 & 0 & \half f^*_{\rho\rho\beta} \\
0 & 0 & -\half f^*_{\rho\rho\beta}
\end{array}
\right)
,\quad
\vvec^*=\left(\begin{array}{r}
0\\
V\\
-V
\end{array}
\right)
\label{M0_v0}
\eea 
where $V=\half\rho^*(A_\rho^*{}^2-B_{\rho\rho}^*)-B^*_\rho$. We want
to deduce from this the expansion of the solution vector,
$\delvec=\delvec^*+r\delvec^*_r+\ldots$  Because $\Mmat^*$ is
degenerate, the zeroth order $\Mmat^*\delvec^*=w\vvec^*$ fixes
the $\delta\beta$-component of $\delvec^*$ as
\be
\delta\beta^* =
w[\rho^*(A_\rho^*{}^2-B_{\rho\rho}^*)-2B^*_\rho]/f^*_{\rho\rho\beta}
\label{crit_pt1}
\ee
and otherwise (only) tells us that the other components must be equal,
$\drc=\drs'\equiv\delta\rho^*$. This is as it should be:
we built the perturbation theory to map the monodisperse
CP to the polydisperse CP, where $\rho_\cl=\rho_\sh'$. Mathematically
it follows from 
the vanishing of the first component of $\vvec^*$, confirming the
intuition that $\delta\lambda_1$ needs to be chosen to make the r.h.s.\
of\eq{dlambda1_eq} zero at the CP.

To determine $\delta\rho^*$, we use the $\order(r)$ condition
$\Mmat^*\delvec^*_r+\Mmat^*_r\delvec^*=w\vvec_r^*$. Multiplying from
the left by $\evec\T=(0,1,1)$ eliminates the first term and gives
$\evec\T\Mmat^*_r\delvec^*=w\evec\T\vvec^*_r$. By inserting the
explicit expressions for $\Mmat^*_r$ and $\vvec^*_r$ from the
small-$r$ expansion and the form
$\delvec^*=(\delta\rho^*,\delta\rho^*,\delta\beta^*)\T$ one then finds
\[
f^*_{\rho\rho\rho\rho}\delta\rho^*
+f^*_{\rho\rho\rho\beta}\delta\beta^* =
w[3A^*_\rho{}^2-3B^*_{\rho\rho}+\rho^*(3A^*_\rho
A^*_{\rho\rho}-B^*_{\rho\rho\rho})]
%\label{crit_pt2}
\]
Together with\eq{crit_pt1} this gives the desired shifts of the
critical density, $\delta\rho^*$, and critical temperature,
$\delta\beta^*$; these results can be checked to agree with the
general polydisperse CP criterion~\cite{Sollich02}, expanded directly
for small $w$. 
Both shifts $\delta\rho^*$ and $\delta\beta^*$ are $\order(w)$, so
that the loci of 
the CPs for increasing $w$ depart from the monodisperse
CP along a line of nontrivial slope in the $(\rho,\beta)$ plane.

Taking the expansion further, we can obtain the slope of the cloud and
shadow curves at the CP, and the location of their maxima. Both
are related to $\delta\beta^*_r=d(\delta\beta)/dr|_{r=0}$, the
$\delta\beta$-component of $\delvec^*_r$. This follows after some
algebra from the first order equation
$\Mmat^*\delvec^*_r+\Mmat^*_r\delvec^*=w\vvec_r^*$ together with the
known $\delvec^*$ as
\be
\delta\beta^*_r = -w\,A^*_\rho
(2A^*_\rho + \rho^* A^*_{\rho\rho})/f^*_{\rho\rho\beta}
\label{delta_beta_r}
\ee
The slope of the cloud curve, plotted as $\beta$ vs $\rho_\cl$, is
$d(\beta+\delta\beta)/d(\rho+\drc)=
[d(\beta+\delta\beta)/dr][d(\rho+\drc)/dr]^{-1}$. At the critical
point, the first factor equals $\delta\beta_r^*$. This is
$\order(w)$, so we only need the $\order(w^0)$ contribution,
$d\rho/dr=1$, of the second factor to find the cloud slope at the
CP as $\delta\beta_r^*$. The shadow slope has the same
modulus but opposite sign since $d\rho'/dr=-1$.
For the cloud and shadow maxima we want to find $r$ such that
$d(\beta+\delta\beta)/dr=0$, giving $br+\delta\beta_r^*=0$ or
$r=-\delta\beta^*_r/b$. Because this 
is $\order(w)$, it is correct to replace $\delta\beta_r$ by
$\delta\beta_r^*$ as we have done. It follows that the cloud and
shadow maxima are to $\order(w)$ located at $\rho_{\cl,\sh}
=\rho^*+\delta\rho^*\mp\delta\beta^*_r/b$, i.e.\ 
equidistant either side of the polydisperse CP.

Finally we turn to purely size-polydisperse
(``scalable''~\cite{Evans01}) systems, in which the interaction
potential remains unchanged when the diameters $\sigma_i$ and position
vectors of all particles are
scaled by a common factor. For pair interactions, this holds if
the dependence on distance $r_{ij}$ of the potential between two particles $i$ and
$j$ can be written as
$v_{ij}(r_{ij})=\Phi(r_{ij}/\ell(\sigma_i,\sigma_j))$ with the
interaction distance $\ell$ a homogeneous function of first degree like
$\ell=(\sigma_i+\sigma_j)/2$ or $\ell=(\sigma_i\sigma_j)^{1/2}$.
This includes e.g.\ polydisperse (additive) hard spheres, or
size-polydisperse Lennard-Jones mixtures.
Simulations and theoretical
calculations~\cite{BelXuBau00,WilSol04,WilFasSol04,FanGazGiaSol06}
show that for such systems, unusually in the presence of
polydispersity, the critical 
point is located very near the maximum of the cloud and shadow curves
(and thus also close to the maximum of a plot of pressure
versus temperature at the onset of phase
coexistence~\cite{BelXuBau00,RasCat03}). Moreover, plotted in terms of
volume fraction instead of density, cloud and shadow curves almost
coincide~\cite{BelXuBau00,WilSol04,WilFasSol04,FanGazGiaSol06}.

This behaviour is fully predicted by our $\order(w)$
perturbation theory. As regards the location of the CP, we need only use that
in size-polydisperse systems~\cite{Evans01}
\be
A=3(\Pi/\rho-1)=3(f_\rho-f/\rho-1)
\label{A}
\ee
Working out $A_\rho$ and $A_{\rho\rho}$,
%It follows that
%$A_\rho=3(f_{\rho\rho}-f_\rho/\rho+f/\rho^2)$ and
%$A_{\rho\rho}=3(f_{\rho\rho\rho}-f_{\rho\rho}/\rho+2f_\rho/\rho^2-2f/\rho^3)$.
substituting into\eq{delta_beta_r} and using
$f^*_{\rho\rho}=f^*_{\rho\rho\rho}=0$ then gives $\delta\beta^*_r=0$.
The cloud and shadow slopes at the CP, $\pm
\delta\beta^*_r$, thus vanish to $\order(w)$, so that the
CP must be at their maximum as claimed.

To see the coincidence of the volume fraction representations of cloud
and shadow, we measure densities in units of $(\pi\sigma_0^3/6)^{-1}$ so
that the volume fraction can be written
$\phi=\int\!d\eps\,\rho(\eps)(1+\eps)^3$. 
Expanding to $\order(w)$
gives for the cloud volume fraction $\phi_\cl=\rho(1+3w)+\drc$, while
for the shadow $\phi_\sh=\rho'(1+3w\Delta A+3w)+\drs'$.
One can now write equations~(\ref{dlambda1_eq}--\ref{Mrow3}) 
in terms of the volume fraction shifts $\dpc=\phi_\cl-\rho$ and
$\dps'=\phi_\sh'-\rho'$.
\iffalse
%
\bea
& &\!\!\!\!\!\!A_\rho\dpc
- A_\rho'\dps' + \Delta A_\beta \,\delta\beta =
\label{delta_lambda1_again}
\\
& &\ \ = 2 w \rho'C'\Delta A + \delta\lambda_1 
-3w\rho' A_\rho'\Delta A + 3w \Delta(\rho A_\rho)
\nonumber\\
%\eea
%and
%\bea
%& &\!\!\!\!\!\!f_{\rho\rho}
%(\dpc-3w\rho)
%+ [f_{\rho\beta} - \Delta f_\beta/\Delta\rho]\, \delta\beta =
%\\
%& &\ \ = w \{\rho'[{\textstyle\half}(\Delta A)^2 +
%\Delta B]/\Delta\rho - \rho B_\rho \}
%\nonumber\\
%& &\!\!\!\!\!\!f_{\rho\rho}' 
%(\dps'-3w\rho'-3w\rho'\Delta A)
%+[f_{\rho\beta}' -\Delta f_\beta/\Delta\rho]\,\delta\beta =
%\\
%& &\ \ = w\{\rho[{\textstyle\half}(\Delta A)^2 + \Delta B]/\Delta\rho
%-\rho' B_\rho' -\rho'A_\rho' \Delta A \}
%\nonumber
%\eea
%%
%or
%
%\bea
& &\!\!\!\!\!\!f_{\rho\rho}
\dpc
+ [f_{\rho\beta} - \Delta f_\beta/\Delta\rho]\, \delta\beta =
\label{Mrow2_phi}\\
& &\ \ = w \{3\rho f_{\rho\rho}+\rho' \Delta B/\Delta\rho - \rho B_\rho
+{\textstyle\half}\rho'(\Delta A)^2/\Delta\rho\}
\nonumber\\
& &\!\!\!\!\!\!f_{\rho\rho}' 
\dps'
+[f_{\rho\beta}' -\Delta f_\beta/\Delta\rho]\,\delta\beta =
\label{Mrow3_phi}
\\
& &\ \ = w\{3\rho' f_{\rho\rho}'+\rho\Delta B/\Delta\rho
-\rho' B_\rho'-{\textstyle\half}\rho(\Delta A)^2/\Delta\rho\}
\nonumber\\
& &\ \ \ \ {}+ w\Delta A[\rho\Delta A/\Delta\rho
+3\rho'f_{\rho\rho}'-\rho'A_\rho']
\nonumber
\eea
%
\fi
Swapping $\rho$
and $\rho'$ gives equations for $\dpc'$ and $\dps$; coincidence requires
that $\dpc=\dps$ and $\dpc'=\dps'$, i.e.\ any point on the
monodisperse binodal is perturbed into cloud and shadow points with
identical volume fractions. The original and swapped equations turn
out to differ by a term proportional to $\rho\Delta A/\Delta\rho
+3\rho'f_{\rho\rho}'-\rho'A_\rho'$ on the r.h.s.; from\eq{A} this
indeed vanishes for size-polydisperse systems.
To ensure that also the temperature shift $\delta\beta$ is the same,
one needs to modify slightly the choice\eq{dlambda1_choice1} of
$\delta\lambda_1$, taking e.g.
\be
\delta\lambda_1=
%-2 w \rho'C'\Delta A + 3w\rho' A_\rho'\Delta A
w\rho'\Delta A(-2 C' + 3 A_\rho')
\label{dlambda1_sizepoly}
\ee
%
%leaves on the r.h.s.\ of\eq{delta_lambda1_again} only the last term,
%which has the correct behaviour under interchange of $\rho$ and
%$\rho'$.
%%; this prescription still gives $\delta\lambda_1=0$ at the
%%%monodisperse
%%CP.

\begin{figure}
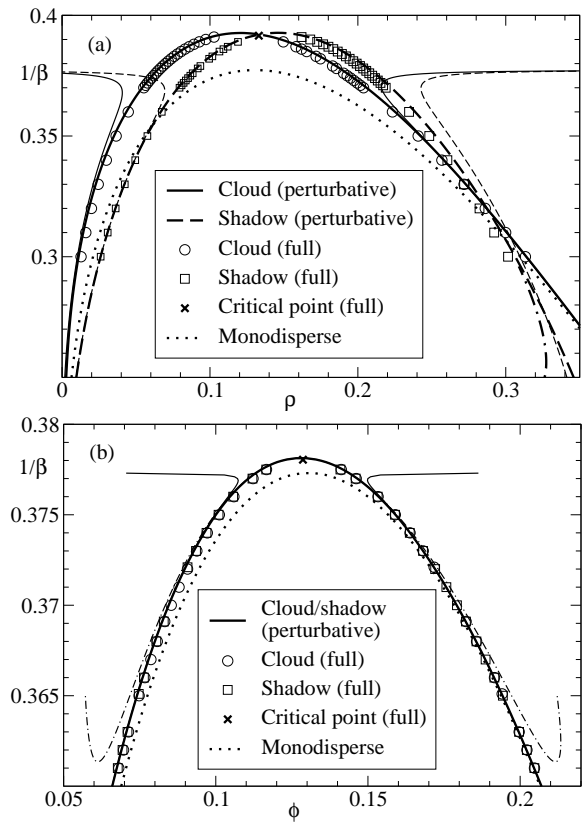

\includegraphics[type=eps,ext=.eps,read=.eps,width=7.5cm,clip=true]{BMCSL_plus_vdW_ampl_s0.07}
\hspace*{-1.1mm}\includegraphics[type=eps,ext=.eps,read=.eps,width=7.62cm,clip=true]{BMCSL_plus_vdW_size_s0.2_reg_and_RMLE_volfracs}
\caption{
  (a) Cloud and shadow curves for a model of a
  Lennard-Jones mixture with polydispersity in both particle diameters and 
  interaction strengths. Symbols:
  full non-perturbative solution. Thick lines: perturbation theory with
  choice\eq{dlambda1_sizepoly}; using\eq{dlambda1_choice1}
  only gives noticeable differences for the high-density part of the
  shadow curve (dash-dotted line).
  Thin lines: constant-$\beta$ approach of~\cite{Evans01}. Dotted line:
  monodisperse binodal. (b)
  Similar plot for a purely size-polydisperse mixture, in the volume
  fraction representation.
%  Cloud and shadow curves coincide, exactly in both perturbative
%  approaches as expected and very
%  nearly for the full solution.
  Dash-dotted line: unregularized perturbation theory.
\label{fig:ampl}
\label{fig:size}
}
\end{figure}
To illustrate the theory we show in Fig.~\ref{fig:ampl}(a) the
predictions for a Lennard-Jones mixture with polydispersity in both
particle diameters and interaction strengths, modelled with a moment
excess free energy~\cite{SolCat98,SolWarCat01} as
in~\cite{WilSolFas05,WilSolFasBuz06}, for a uniform (``top hat'') size
distribution $p(\eps)$ of width $w^{1/2}=0.07$. The generalized
perturbation theory produces cloud and shadow curves with smoothly
connecting sub- and super-critical branches, avoiding the divergences
near the monodisperse CP of the constant-$\beta$
approach~\cite{Evans01}. Comparing with the symbols, our perturbation
theory is quantitatively closer to the full solution of the
polydisperse phase equilibrium conditions also elsewhere, in
particular for the high-density branch of the shadow curve. The
overall level of agreement is quite remarkable given that the shifts from the
monodisperse reference system (dotted line) are substantial.

Fig.~\ref{fig:size}(b) shows similar results for a mixture with pure size
polydispersity~\cite{WilSol04,WilFasSol04}.  In the volume fraction
representation cloud and shadow curves coincide as predicted, and the
CP is very near their maximum. Again, our approach avoids
divergences in this region and is quantitatively accurate even for the
substantial degree of polydispersity $w^{1/2}=0.2$ considered here,
i.e.\ a size standard that is $20\%$ of the mean. The
dash-dotted line gives the raw results of the
perturbation theory; the unphysical upward curvature comes from an
increase in the temperature shifts $|\delta\beta|$. This illustrates
that, away from the CP where we should have
$\delta\lambda_1=0$, it is not obvious how to choose $\delta\lambda_1$
to extract the most reliable predictions for finite $w$. Intuitively,
one expects temperature shifts to be needed primarily in the
critical region and to become smaller elsewhere. This can be enforced by
e.g.\ taking the raw $\delta\beta$ and using a soft thresholding
function so that values with $|\delta\beta|>|\delta\beta^*|$ are
reassigned to be no larger than $2 |\delta\beta^*|$; $\drc$ and $\drs'$
are then recalculated from~(\ref{Mrow2},\ref{Mrow3}). The thick line
in Fig.~\ref{fig:size} shows that this gives accurate and physically
sensible results.
% Whether more principled methods for such
%regularization can be found, e.g.\ by choosing $\delta\beta$ to
%minimize the neglected $\order(w^2)$ corrections, is an open question.

The
% perturbation theory
approach presented here allows the calculation of full
phase diagrams for generic weakly polydisperse systems, giving access
to key features such as the separation of the maxima of cloud and
shadow from the CP, the differing cloud and shadow slopes at the CP,
and the polydispersity-induced shifts of the critical parameters.
Even for excess free energies with
moment structure it can be helpful in avoiding the
computational complexities of a full phase equilibrium
calculation~\cite{Sollich02}, while remaining quantitatively accurate
in a substantial range of polydispersities (up to $20\%$ in
Fig.~\ref{fig:size}) that includes typical experimental
values~\cite{PusVan86}.
The theory also generalizes easily to
phase equilibria inside the coexistence region, as well as systems
with several polydisperse attributes.

\bibliographystyle{apsrev}
\bibliography{/home/psollich/references/references.bib}

\begin{thebibliography}{14}
\expandafter\ifx\csname natexlab\endcsname\relax\def\natexlab#1{#1}\fi
\expandafter\ifx\csname bibnamefont\endcsname\relax
  \def\bibnamefont#1{#1}\fi
\expandafter\ifx\csname bibfnamefont\endcsname\relax
  \def\bibfnamefont#1{#1}\fi
\expandafter\ifx\csname citenamefont\endcsname\relax
  \def\citenamefont#1{#1}\fi
\expandafter\ifx\csname url\endcsname\relax
  \def\url#1{\texttt{#1}}\fi
\expandafter\ifx\csname urlprefix\endcsname\relax\def\urlprefix{URL }\fi
\providecommand{\bibinfo}[2]{#2}
\providecommand{\eprint}[2][]{\url{#2}}

\bibitem[{\citenamefont{Pusey and {van Megen}}(1986)}]{PusVan86}
\bibinfo{author}{\bibfnamefont{P.~N.} \bibnamefont{Pusey}} \bibnamefont{and}
  \bibinfo{author}{\bibfnamefont{W.}~\bibnamefont{{van Megen}}},
  \bibinfo{journal}{Nature} \textbf{\bibinfo{volume}{320}},
  \bibinfo{pages}{340} (\bibinfo{year}{1986}).

\bibitem[{\citenamefont{Evans et~al.}(1998)\citenamefont{Evans, Fairhurst, and
  Poon}}]{EvaFaiPoo98}
\bibinfo{author}{\bibfnamefont{R.~M.~L.} \bibnamefont{Evans}},
  \bibinfo{author}{\bibfnamefont{D.~J.} \bibnamefont{Fairhurst}},
  \bibnamefont{and} \bibinfo{author}{\bibfnamefont{W.~C.~K.}
  \bibnamefont{Poon}}, \bibinfo{journal}{Phys.\ Rev.\ Lett.}
  \textbf{\bibinfo{volume}{81}}, \bibinfo{pages}{1326} (\bibinfo{year}{1998}).

\bibitem[{\citenamefont{Evans}(2001)}]{Evans01}
\bibinfo{author}{\bibfnamefont{R.~M.~L.} \bibnamefont{Evans}},
  \bibinfo{journal}{J.\ Chem.\ Phys.} \textbf{\bibinfo{volume}{114}},
  \bibinfo{pages}{1915} (\bibinfo{year}{2001}).

\bibitem[{\citenamefont{Rasc{\'{o}}n and Cates}(2003)}]{RasCat03}
\bibinfo{author}{\bibfnamefont{C.}~\bibnamefont{Rasc{\'{o}}n}}
  \bibnamefont{and} \bibinfo{author}{\bibfnamefont{M.~E.} \bibnamefont{Cates}},
  \bibinfo{journal}{J.\ Chem.\ Phys.} \textbf{\bibinfo{volume}{118}},
  \bibinfo{pages}{4312} (\bibinfo{year}{2003}).

\bibitem[{\citenamefont{Wilding and Sollich}(2004)}]{WilSol04}
\bibinfo{author}{\bibfnamefont{N.~B.} \bibnamefont{Wilding}} \bibnamefont{and}
  \bibinfo{author}{\bibfnamefont{P.}~\bibnamefont{Sollich}},
  \bibinfo{journal}{Europhys.\ Lett.} \textbf{\bibinfo{volume}{67}},
  \bibinfo{pages}{219} (\bibinfo{year}{2004}).

\bibitem[{\citenamefont{Wilding et~al.}(2004)\citenamefont{Wilding, Fasolo, and
  Sollich}}]{WilFasSol04}
\bibinfo{author}{\bibfnamefont{N.~B.} \bibnamefont{Wilding}},
  \bibinfo{author}{\bibfnamefont{M.}~\bibnamefont{Fasolo}}, \bibnamefont{and}
  \bibinfo{author}{\bibfnamefont{P.}~\bibnamefont{Sollich}},
  \bibinfo{journal}{J.\ Chem.\ Phys.} \textbf{\bibinfo{volume}{121}},
  \bibinfo{pages}{6887} (\bibinfo{year}{2004}).

\bibitem[{\citenamefont{Sollich and Cates}(1998)}]{SolCat98}
\bibinfo{author}{\bibfnamefont{P.}~\bibnamefont{Sollich}} \bibnamefont{and}
  \bibinfo{author}{\bibfnamefont{M.~E.} \bibnamefont{Cates}},
  \bibinfo{journal}{Phys.\ Rev.\ Lett.} \textbf{\bibinfo{volume}{80}},
  \bibinfo{pages}{1365} (\bibinfo{year}{1998}).

\bibitem[{\citenamefont{Sollich et~al.}(2001)\citenamefont{Sollich, Warren, and
  Cates}}]{SolWarCat01}
\bibinfo{author}{\bibfnamefont{P.}~\bibnamefont{Sollich}},
  \bibinfo{author}{\bibfnamefont{P.~B.} \bibnamefont{Warren}},
  \bibnamefont{and} \bibinfo{author}{\bibfnamefont{M.~E.} \bibnamefont{Cates}},
  \bibinfo{journal}{Adv.\ Chem.\ Phys.} \textbf{\bibinfo{volume}{116}},
  \bibinfo{pages}{265} (\bibinfo{year}{2001}).

\bibitem[{\citenamefont{Sollich}(2002)}]{Sollich02}
\bibinfo{author}{\bibfnamefont{P.}~\bibnamefont{Sollich}},
  \bibinfo{journal}{J.\ Phys.\ Cond.\ Matt.} \textbf{\bibinfo{volume}{14}},
  \bibinfo{pages}{R79} (\bibinfo{year}{2002}).

\bibitem[{\citenamefont{Kita et~al.}(1997)\citenamefont{Kita, Dobashi,
  Yamamoto, Nakata, and Kamide}}]{KitDobYamNakKam97}
\bibinfo{author}{\bibfnamefont{R.}~\bibnamefont{Kita}},
  \bibinfo{author}{\bibfnamefont{T.}~\bibnamefont{Dobashi}},
  \bibinfo{author}{\bibfnamefont{T.}~\bibnamefont{Yamamoto}},
  \bibinfo{author}{\bibfnamefont{M.}~\bibnamefont{Nakata}}, \bibnamefont{and}
  \bibinfo{author}{\bibfnamefont{K.}~\bibnamefont{Kamide}},
  \bibinfo{journal}{Phys.\ Rev.\ E} \textbf{\bibinfo{volume}{55}},
  \bibinfo{pages}{3159} (\bibinfo{year}{1997}).

\bibitem[{\citenamefont{Bellier-Castella
  et~al.}(2000)\citenamefont{Bellier-Castella, Xu, and Baus}}]{BelXuBau00}
\bibinfo{author}{\bibfnamefont{L.}~\bibnamefont{Bellier-Castella}},
  \bibinfo{author}{\bibfnamefont{H.}~\bibnamefont{Xu}}, \bibnamefont{and}
  \bibinfo{author}{\bibfnamefont{M.}~\bibnamefont{Baus}}, \bibinfo{journal}{J.\
  Chem.\ Phys.} \textbf{\bibinfo{volume}{113}}, \bibinfo{pages}{8337}
  (\bibinfo{year}{2000}).

\bibitem[{\citenamefont{Fantoni et~al.}(2006)\citenamefont{Fantoni, Gazzillo,
  Giacometti, and Sollich}}]{FanGazGiaSol06}
\bibinfo{author}{\bibfnamefont{R.}~\bibnamefont{Fantoni}},
  \bibinfo{author}{\bibfnamefont{D.}~\bibnamefont{Gazzillo}},
  \bibinfo{author}{\bibfnamefont{A.}~\bibnamefont{Giacometti}},
  \bibnamefont{and} \bibinfo{author}{\bibfnamefont{P.}~\bibnamefont{Sollich}},
  \bibinfo{journal}{J.\ Chem.\ Phys.} \textbf{\bibinfo{volume}{125}},
  \bibinfo{pages}{164504} (\bibinfo{year}{2006}).

\bibitem[{\citenamefont{Wilding et~al.}(2005)\citenamefont{Wilding, Sollich,
  and Fasolo}}]{WilSolFas05}
\bibinfo{author}{\bibfnamefont{N.~B.} \bibnamefont{Wilding}},
  \bibinfo{author}{\bibfnamefont{P.}~\bibnamefont{Sollich}}, \bibnamefont{and}
  \bibinfo{author}{\bibfnamefont{M.}~\bibnamefont{Fasolo}},
  \bibinfo{journal}{Phys.\ Rev.\ Lett.} \textbf{\bibinfo{volume}{95}},
  \bibinfo{pages}{155701} (\bibinfo{year}{2005}).

\bibitem[{\citenamefont{Wilding et~al.}(2006)\citenamefont{Wilding, Sollich,
  Fasolo, and Buzzacchi}}]{WilSolFasBuz06}
\bibinfo{author}{\bibfnamefont{N.~B.} \bibnamefont{Wilding}},
  \bibinfo{author}{\bibfnamefont{P.}~\bibnamefont{Sollich}},
  \bibinfo{author}{\bibfnamefont{M.}~\bibnamefont{Fasolo}}, \bibnamefont{and}
  \bibinfo{author}{\bibfnamefont{M.}~\bibnamefont{Buzzacchi}},
  \bibinfo{journal}{J.\ Chem.\ Phys.} \textbf{\bibinfo{volume}{125}},
  \bibinfo{pages}{014908} (\bibinfo{year}{2006}).

\end{thebibliography}

\end{document}